\documentclass[runningheads]{svmult}
\usepackage{makeidx}   
\usepackage{graphicx}  
\usepackage{subeqnar}  
\usepackage{multicol}  
\usepackage{cropmark} 
\usepackage{physprbb}  
\newcommand{\lsim}{\mathrel{\mathop{\kern 0pt \rlap
  {\raise.2ex\hbox{$<$}}}
  \lower.9ex\hbox{\kern-.190em $\sim$}}}
\newcommand{\gsim}{\mathrel{\mathop{\kern 0pt \rlap
  {\raise.2ex\hbox{$>$}}}
  \lower.9ex\hbox{\kern-.190em $\sim$}}}

\begin{document}
\title*{Neutrino Oscillation Effects in Indirect Detection of Dark Matter}
\toctitle{Neutrino Oscillation Effects in Indirect Detection of Dark Matter}
%
%
\titlerunning{Neutrino Oscillation Effects in Indirect Detection of Dark Matter}
%
\author{Nicolao Fornengo}
\authorrunning{Nicolao Fornengo}
%
%
\institute{Dipartimento di Fisica Teorica, Universit\`a di Torino \\
     and Istituto Nazionale di Fisica Nucleare \\
     via P. Giuria 1, I-10125 Torino, Italy}

\maketitle              

\begin{abstract}
  If neutrino oscillation plays a role in explaining the atmospheric
  neutrino deficit, then the same phenomenon would necessarily affect
  also the dark matter indirect-detection signal which consists in a
  muon-neutrino flux produced by neutralino annihilation in the Earth
  core. In this paper we investigate to which extent the upgoing-muon
  signal originated by neutralinos captured inside the Earth would be
  affected by the presence of $\nu_\mu \rightarrow \nu_\tau$
  oscillation.
\end{abstract}

\section{Introduction}
Among the different techniques which have been proposed to search for
dark matter particles \cite{nota0}, detection of a neutrino flux by
means of neutrino telescopes represents certainly an interesting tool.
Although direct detection \cite{ICTP,JKG,old} at present appears to be
somewhat more sensitive to neutralino dark matter
 \cite{probing,further,notadir}, nevertheless all the different
possibilities are worth being explored.  In this paper we discuss the
flux of upgoing muons which are a consequence of $\nu_\mu$'s produced
inside the Earth, with special emphasis on the role which is played in
this kind of searches by the possible presence of neutrino
oscillation.

\section{Upgoing $\mu$'s from neutralino annihilation in the Earth}
Neutralinos can be gravitationally captured inside astrophysical
bodies \cite{GPS}, like the Earth and the Sun. Their subsequent
annihilation can produce a flux of neutrinos, which then travel
toward a detector located underground below the Earth surface.
The differential flux, for each neutrino flavour $i$, is defined as
\begin{equation}
\Phi^0_{\stackrel{(-)}{\nu_i}} (E_\nu) \equiv 
\frac{dN_{\stackrel{(-)}{\nu_i}}}{dE_\nu} =
\frac{\Gamma_A}{4\pi d^2} \sum_{F,f}
B^{(F)}_{\chi f}\frac{dN_{f {\stackrel{(-)}{\nu_i}}}}{dE_i} \, ,  
\label{eq:fluxnu}
\end{equation}
where $\Gamma_A$ denotes the neutralino annihilation rate, $d$ is the
distance between the detector and the source (which can be the center
of the Earth or the Sun), $F$ is an index which lists all the
possibile final states which can be produced by neutralino
pair-annihilation, $B^{(F)}_{\chi f}$ denotes, for each final state
$F$, the branching ratios into heavy quarks, $\tau$ leptons and
gluons. The differential spectra of neutrinos and antineutrinos
generated by the $\tau$ and by hadronization of quarks and gluons and
the subsequent semileptonic decays of the produced hadrons are denoted
by $dN_{f {\stackrel{(-)}{\nu_i}} }/dE_{\nu}$. For more details, see
for instance Refs. \cite{ICTP,JKG,noi_nuflux,altri_nuflux}. Here we
only recall that the annihilation rate depends, through its relation
with the capture rate of neutralinos in the Earth, on some
astrophysical parameters, the most relevant of which is the local
density $\rho_l$.

The best way of identifying the presence of these fluxes relies on the
possibility to detect upward going muons inside a neutrino telescope.
These upgoing muons would be produced by the $\nu_\mu$ component of
the neutrino fluxes of Eq.(1). The charged-current interaction of the
$\nu_\mu$'s with the rock below and close to the detector would
produce a flux of muons. A double-differential muon flux is defined as
\begin{equation}
\frac{d^2 N_\mu}{d E_\mu d E_\nu} = \sum_j
N_A \int_0^\infty dX \int_{E_\mu}^{E_\nu} d E'_\mu 
g(E_\mu, E'_\mu; X) \Phi^0_j(E_\nu) \frac{d \sigma_j(E_\nu,E'_\mu)}{d
  E'_\mu}\, ,
\end{equation}
where $j = \nu_\mu, \bar\nu_\mu$, $N_A$ is the Avogadro's number,
$g(E_\mu,E'_\mu; X)$ is the survival probability that a muon of
initial energy $E'_\mu$ will have a final energy $E_\mu$ after
propagating along a distance $X$ inside the rock and $d
\sigma_j (E_\nu,E'_\mu) / d E'_\mu$ is the
charged-current cross section for the production of a muon of energy $
E'_\mu$ from a neutrino (antineutrino) of energy $E_\nu$.

A useful quantity for our discussion is
the muon response function
\begin{equation}
\frac{d N_\mu}{d E_\nu} = \int_{E^{\mathrm{th}}}^{E_\nu}
d E_\mu \; \frac{d^2 N_\mu}{d E_\mu d E_\nu}\, ,
\end{equation}
where $E^{\mathrm{th}}$ is threshold energy for detection of up--going
muons. For Super Kamiokande and MACRO, $E^{\mathrm{th}} \simeq 1.5$
GeV \cite{oscill_exp}.  The muon response function identifies the
neutrino energy range that is mostly responsile for the up--going muon
signal.  Fig. 1 shows a few examples of it, plotted as functions of
the variable $x = E_\nu/m_\chi$, where $m_\chi$ denotes the neutralino
mass. Fig. 1 shows an approximate scaling of $dN_\mu/dE_\nu$ with
$m_\chi$. The maximum of the muon response occurs for neutrino
energies of about $E_\nu \simeq (0.4 - 0.6) \; m_\chi$, with a half
width which extends from $E_\nu \simeq 0.1\; m_\chi$ to $E_\nu \simeq
0.8 \; m_\chi$.

The quantity which is actually measured is the total flux of up--going
muons, which is defined as
\begin{equation}
\Phi_\mu = \int_{E^{\mathrm{th}}}^{m_\chi}
d E_\nu \; \frac{d N_\mu}{d E_\nu}\, .
\end{equation}
$\Phi_\mu$ can be calculated once a specific supersymmetric model is
adopted. In the case of a model where all the supersymmetric
parameters are defined and set at the electroweak scale (which we call
here MSSM), the result for $\Phi_\mu$ is shown in Fig. 2. We have
varied the MSSM parameters in the ranges: $20\;\mbox{GeV} \leq M_2
\leq 1000\;\mbox{GeV}$, $20\;\mbox{GeV} \leq |\mu| \leq
1000\;\mbox{GeV}$, $90\;\mbox{GeV} \leq m_A \leq 1000\;\mbox{GeV}$,
$100\;\mbox{GeV} \leq m_0 \leq 1000\;\mbox{GeV}$, $-3 \leq {\rm A}
\leq +3,\; 1 \leq \tan \beta \leq 50$. Up-to-date bounds and
limits coming from accelerators and from BR($b\rightarrow s \gamma)$
have been imposed.  For a definition of supersymmetric models and
their parameters, as well as the implementation of the experimental
limits on susy searches, see Ref. \cite{probing}. For calculations of
$\Phi_\mu$ in supergravity inspired (SUGRA) models, see Ref.
\cite{nusugra}.

Fig. 2 also shows the present most stringent upper limit obtained by
the MACRO Collaboration \cite{MACRO}. Super Kamiokande recently also 
reported a similar upper bound \cite{SK_WIMP}.

\section{Neutrino oscillation effect on the up--going muon signal}
The atmospheric neutrino deficit strongly points toward the indication
that the $\nu_\mu$ may oscillate. The oscillation channel which best
describes the anomaly is $\nu_\mu \rightarrow \nu_\tau$ vacuum
oscillation \cite{FGGV,SK_OSC_nos}. If this is the case, also the
$\nu_\mu$ produced by neutralino annihilations would undergo an
oscillation process. The range of energies involved in both
atmospheric and neutralino--produced neutrinos is approximately the
same, while the baseline of oscillation of the two neutrino components
is different
Atmospheric neutrinos which induce upgoing muons cover a range of
pathlengths which ranges from twice the Earth's radius, for vertical
muons, to much shorter distances in the case of horizonthal muons.  On
the contrary, neutrinos produced by neutralino annihilation in the central
part of the Earth travel a distance of the order of the Earth's radius
to reach the detector. On the basis of the features of the $\nu_\mu$
oscillation which are required to fit the experimental data on
atmospheric neutrinos \cite{FGGV,SK_OSC_nos}, we expect that also the
neutrino flux from dark matter annihilation would be affected
\cite{Ellis}.

For $\nu_\mu \rightarrow \nu_\tau$ oscillation, the $\nu_\mu$ flux is
reduced because of oscillation, but we have to take into account
that neutralino annihilation can also produce $\nu_\tau$ which in turn can
oscillate into $\nu_\mu$ and contribute to the up--going muon flux.
The muon neutrino flux can therefore be expressed as (we are
considering only two-flavour oscillation)
\begin{equation}
\Phi_{{\stackrel{(-)}{\nu_\mu}}} (E_\nu) =
\Phi^0_{{\stackrel{(-)}{\nu_\mu}}}\; 
P^{\mathrm{vac}} ({{\stackrel{(-)}{\nu_\mu}}} \rightarrow
{{\stackrel{(-)}{\nu_\mu}}})
 +  
\Phi^0_{{\stackrel{(-)}{\nu_\tau}}}\; 
[1-P^{\mathrm{vac}} ({{\stackrel{(-)}{\nu_\mu}}} \rightarrow
{{\stackrel{(-)}{\nu_\mu}}}) ] \, ,
\end{equation}
where the vacuum survival probability is \cite{Kim}
\begin{equation}
P^{\mathrm{vac}} ({{\stackrel{(-)}{\nu_\mu}}} \rightarrow
{{\stackrel{(-)}{\nu_\mu}}})
 =  
 1 - \sin^2(2\theta)\sin^2
\left (
\frac{1.27 \Delta m^2 (\mathrm{eV}^2) R(\mathrm{Km})}
{E_\nu (\mathrm{GeV})}
\right ) \, ,
\label{vac}
\end{equation}
where $\Delta m^2$ is the mass square difference of the two neutrino
mass eigenstates, $\theta$ is the mixing angle in vacuum and $R$ is
the Earth's radius.  Fig. 3 shows the survival probability for two
different values of the neutrino oscillation parameters which are
inside their allowed ranges \cite{FGGV,SK_OSC_nos}. Smaller (larger)
values of $\Delta m^2$ have the effect of shifting the curves to the
left (right).  Comparing Fig. 1 with Fig. 3, we notice that the
reduction of the up--going muon flux is stronger when there is
matching between the the energy $E_\nu^1 \simeq 5.2 \cdot 10^{3}
\Delta m^2 (\mathrm{eV}^2)$ of the first (from the right) minimum of
the survival probability and the energy $E_\nu \simeq 0.5 m_\chi$
which is responsible for most of the muon response in the detector.
This implies that a maximum reduction of the signal could occur for
neutralino masses of the order of $m_\chi (\mathrm{GeV}) \simeq 10^4
\Delta m^2 (\mathrm{eV}^2)$.  The $\nu_\tau \rightarrow \nu_\mu$
oscillation makes the reduction of the muon flux less severe, but it
is not able to completely balance the reduction effect because the
original $\nu_\tau$ flux at the source is smaller than the $\nu_\mu$
flux. Therefore, the overall effect of the neutrino oscillation is to
reduce the up--going muon signal. The upgoing muon flux for a
neutralino in the MSSM, when neutrino oscillation is included, is
given in Fig. 4.  This, when compared with Fig. 2, shows the effect
induced by the presence of oscillation. The ratio of the up--going
muon signals in the presence and in the absence of oscillation is
plotted in fig. 5.  We notice that the strongest effect occurs for
light neutralinos, since in this case the muon flux is mostly produced
from neutrinos whose energy is in the range of maximal suppression for
the oscillation phenomenon. The effect is between 0.5 and 0.8 for
$m_\chi \lsim 100$ GeV.  On the contrary, the fluxes for larger masses
are less affected, and the reduction is less than about 20\% for
$m_\chi \gsim 200$ GeV. Figs. 2, 4 and 5 update the corresponding
figures of Ref.\cite{meprev} by the inclusion of the new limits from
accelerator on Higgs and Supersymmetry searches.

\section{Conclusions}
In this paper we have discussed to which extent neutrino oscillation
can affect the up--going muon signal from neutralino annihilation in
the Earth. While the experimental upper limit is, at present,
practically not affected by neutrino oscillation
\cite{MACRO}, the theoretical predictions are reduced in the presence
of oscillation. By adopting the neutrino oscillation parameters
deduced {}from the fits on the atmospheric neutrino data
\cite{FGGV,SK_OSC_nos}, the effect is always larger for lighter
neutralinos. For $\nu_\mu \rightarrow \nu_\tau$ the reduction is between
0.5 and 0.8 for $m_\chi \lsim 100$ GeV and less than about 20\% for
$m_\chi \gsim 200$ GeV.

\section*{Acknowledgments}
This work was partially supported 
by the Research Grants of the Italian Ministero
dell'Universit\`a e della Ricerca Scientifica e Tecnologica 
(MURST) within the {\sl Astroparticle Physics Project}.

\begin{figure}[b]
\begin{center}

  \includegraphics[width=.8\textwidth,bbllx=50bp,bblly=200bp,bburx=520bp,bbury=650bp,clip=]{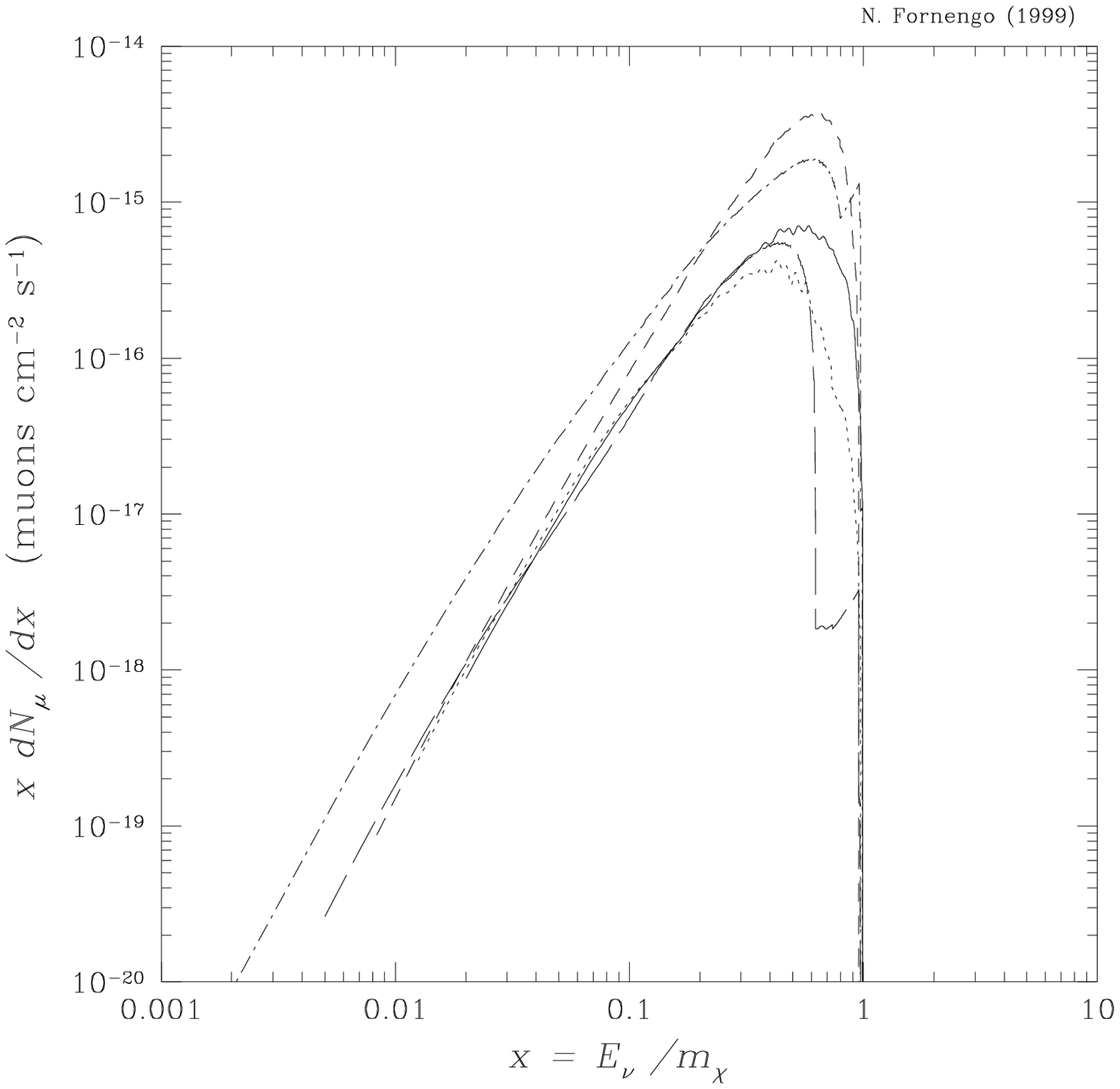}
\end{center}
\caption[]{Muon response function $d N_\mu / d\log x$ vs.
  the parent-neutrino fractional energy $x = E_\nu / m_\chi$ for
  neutralino annihilation in the Earth. Different curves refer to
  different neutralino masses : $m_\chi = 50$ GeV (solid), $m_\chi =
  80$ GeV (dotted), $m_\chi = 120$ GeV (shot--dashed), $m_\chi = 200$
  GeV (long--dash), $m_\chi = 500$ GeV (dot--dashed).}
\end{figure}

\begin{figure}[b]
\begin{center}
\includegraphics[width=.8\textwidth,bbllx=50bp,bblly=200bp,bburx=520bp,bbury=650bp,clip=]{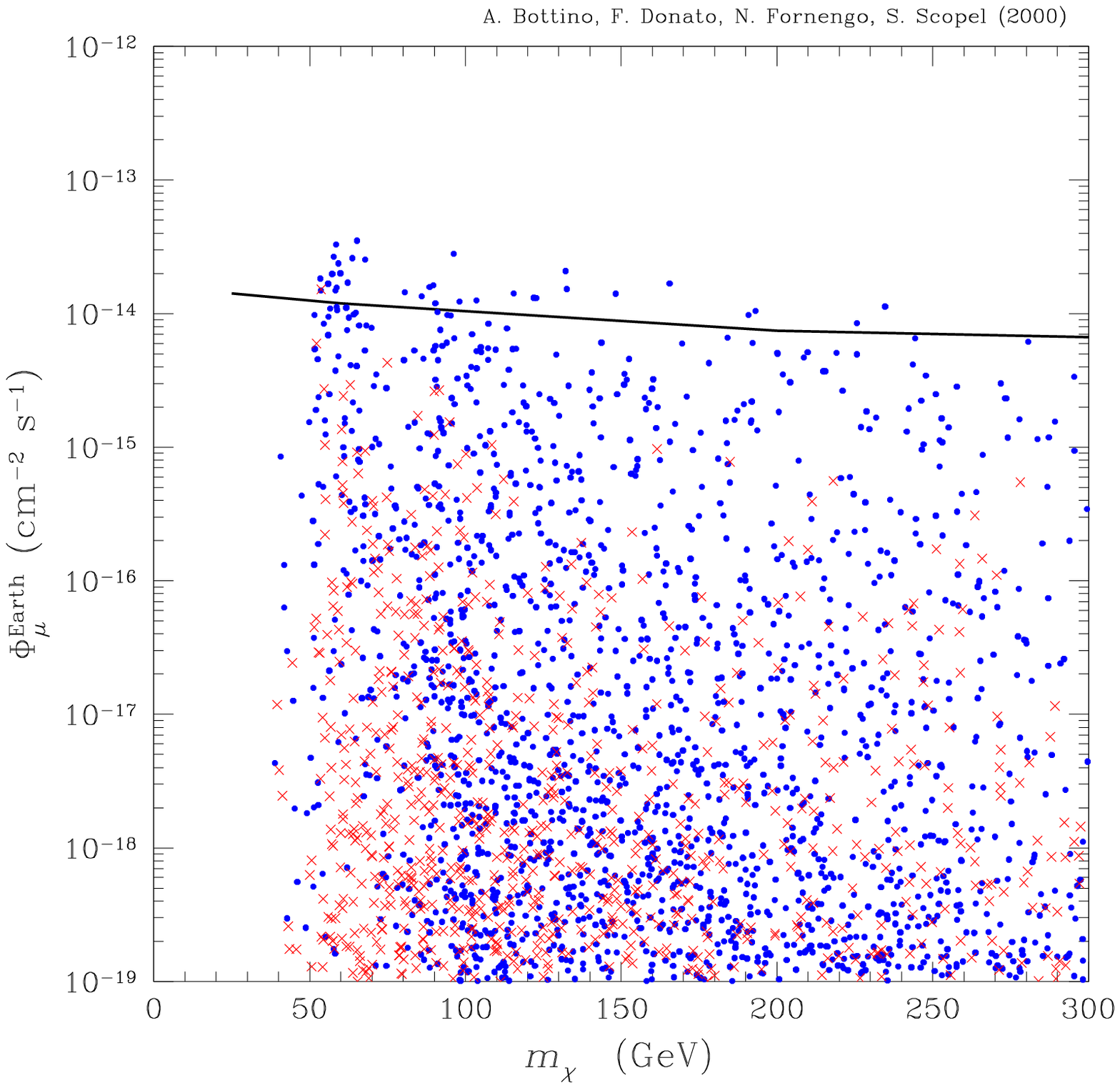}
\end{center}
\caption[]{Flux of up--going muons $\Phi_\mu^{\mathrm{Earth}}$ from
  neutralino annihilation in the Earth, plotted as a function of
  $m_\chi$. The solid line denotes the present upper limit from MACRO
   \cite{MACRO}.  Crosses denote supersymmetric configurations for
  which the neutralino relic abundance $\Omega_\chi h^2$ is larger
  than, or equal to, the value 0.05 (but not in excess of its
  cosmological upper bound of 0.7 \cite{probing}). Dots stand for $\Omega_\chi h^2<
  0.05$.  The $\nu_\mu$'s are assumed not to oscillate.}
\end{figure}

\begin{figure}[b]
\begin{center}
  \includegraphics[width=.8\textwidth,bbllx=50bp,bblly=200bp,bburx=520bp,bbury=650bp,clip=]{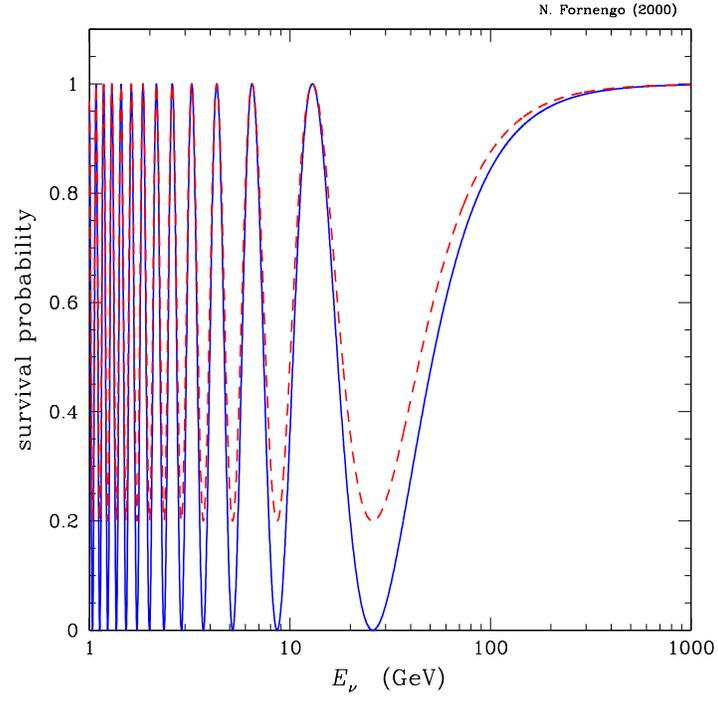}
\end{center}
\caption[]{$\nu_\mu$ survival probability in the case of
  $\nu_\mu \rightarrow \nu_\tau$ oscillation. The solid line refers to
  $\sin^2 (2\theta) = 1$, the dashed line is for $\sin^2 (2\theta)
  = 0.8$.  In both cases, $\Delta m^2 = 5\cdot 10^{-3}$ eV $^{-2}$.}
\end{figure}

\begin{figure}[b]
\begin{center}
\includegraphics[width=.8\textwidth,bbllx=50bp,bblly=200bp,bburx=520bp,bbury=650bp,clip=]{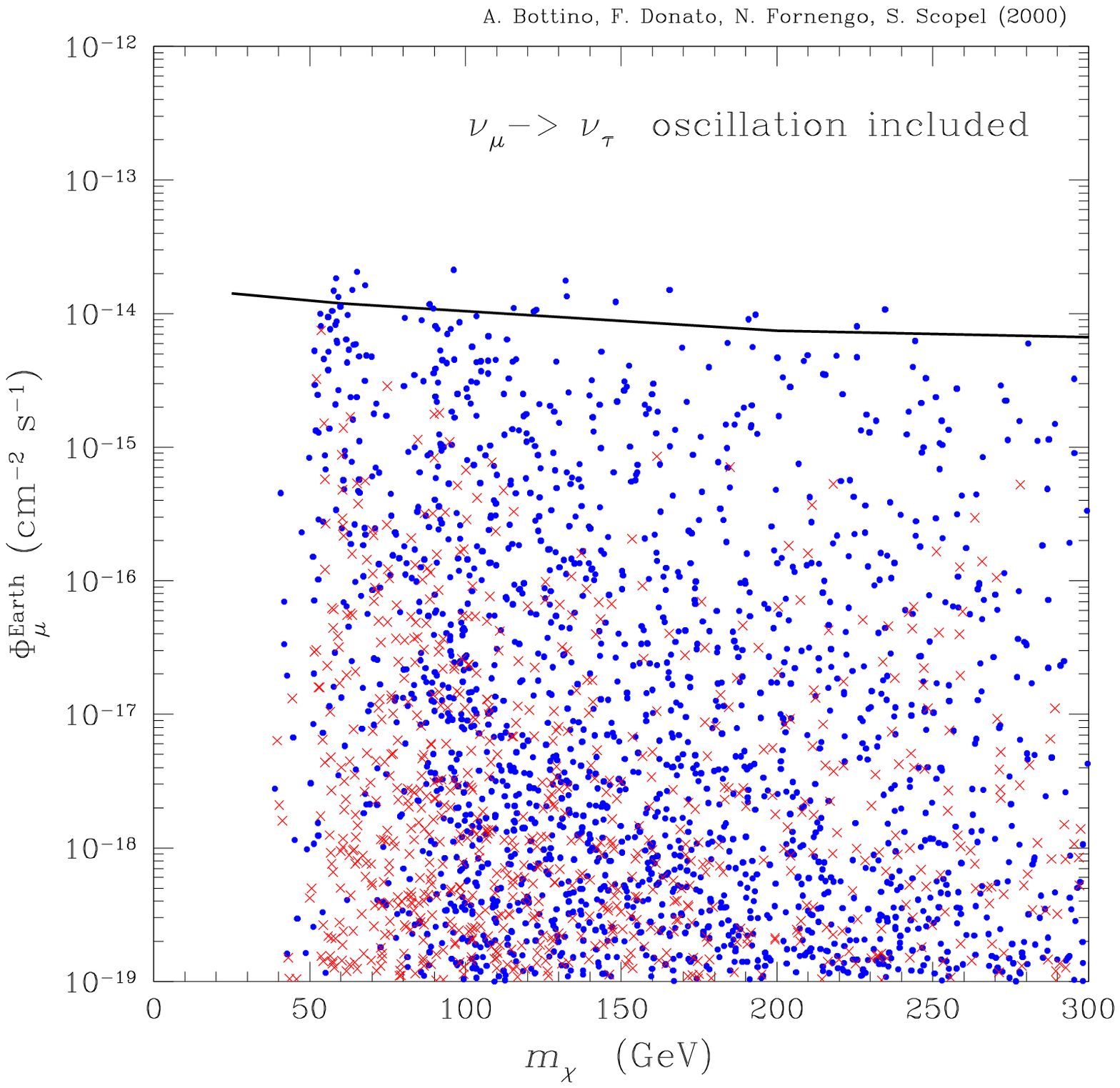}
\end{center}
\caption[]{Flux of up--going muons $\Phi_\mu^{\mathrm{Earth}}$ from
  neutralino annihilation in the Earth, plotted as a function of
  $m_\chi$. The solid line denotes the present upper limit from MACRO
   \cite{MACRO}.  Crosses denote supersymmetric configurations for
  which the neutralino relic abundance $\Omega_\chi h^2$ is larger
  than, or equal to, the value 0.05 (but not in excess of its
  cosmological upper bound of 0.7 \cite{probing}). Dots stand for $\Omega_\chi h^2<
  0.05$.  The $\nu_\mu$'s are assumed to oscillate into $\nu_\tau$'s,
  with oscillation parameters fixed at the best-fit values of Ref.
  \cite{FGGV}: $\sin^2(2 \theta) = 1$ and $\Delta m^2 = 3 \cdot
  10^{-3}$ eV$^2$.}
\end{figure}

\begin{figure}[b]
\begin{center}
\includegraphics[width=.8\textwidth,bbllx=50bp,bblly=200bp,bburx=520bp,bbury=650bp,clip=]{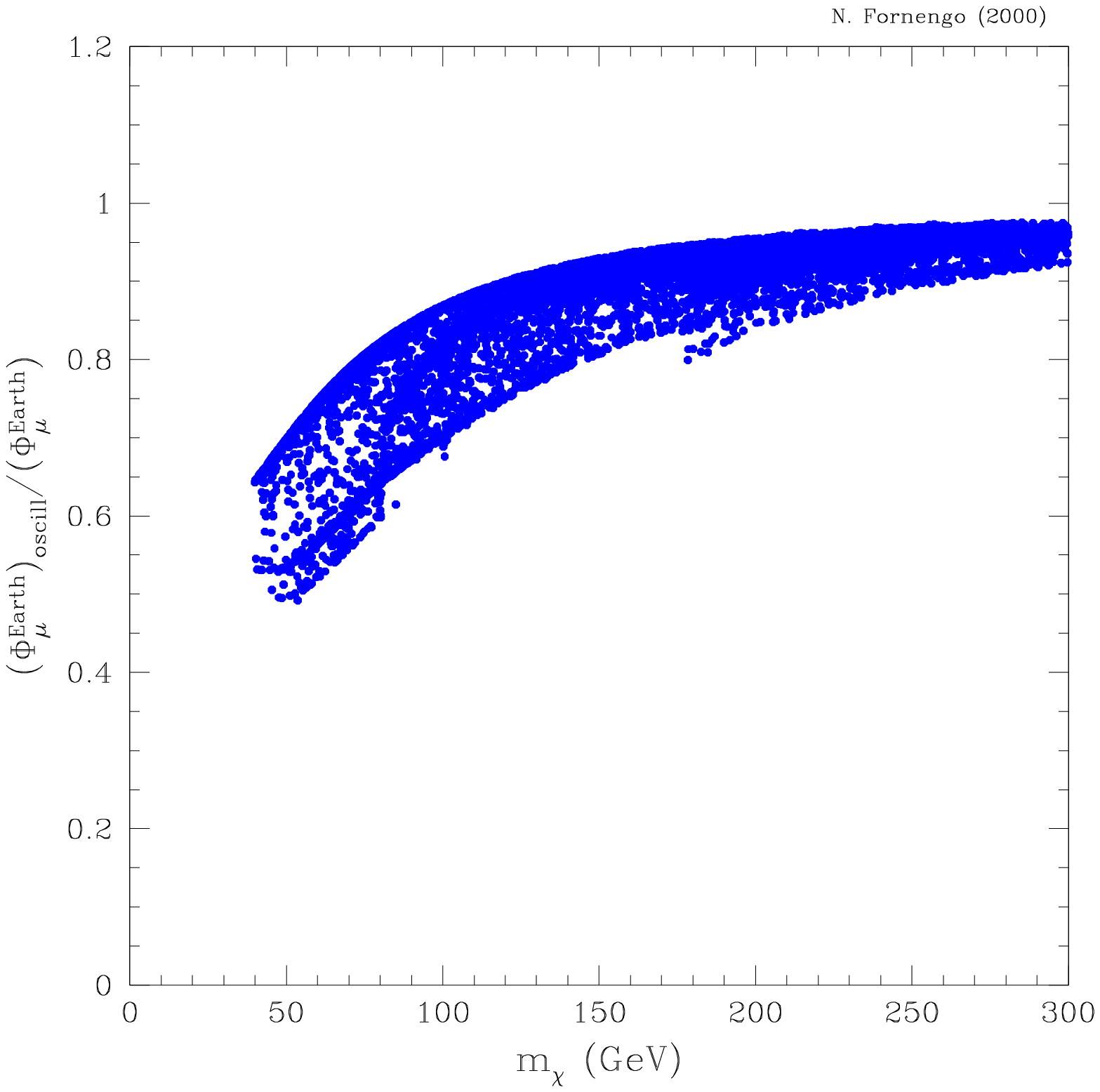}
\end{center}
\caption[]{Scatter plot of the ratio 
  $(\Phi_\mu^{\rm Earth})_{\rm oscill}/\Phi_\mu^{\rm Earth}$ vs. the
  neutralino mass $m_\chi$. $(\Phi_\mu^{\rm Earth})_{\rm oscill}$ is
  the up--going muon flux in the case of $\nu_\mu \rightarrow
  \nu_\tau$ oscillation (shown in Fig.4), while $\Phi_\mu^{\rm Earth}$
  is the corresponding flux in the case of no oscillation and (plotted
  in Fig.2). For the oscillation case, the neutrino parameters have
  been set at the best-fit values of Ref. \cite{FGGV}: $\sin^2
  (2\theta) = 1$ and $\Delta m^2 = 3\cdot 10^{-3}$ eV $^{-2}$.}
\end{figure}

%


\begin{thebibliography}{8.}
\addcontentsline{toc}{section}{References}

\bibitem{nota0} For an introduction to dark matter and to neutralino
  searches, as well as for a more comprehensive list of references,
  see for instance, Ref. \cite{ICTP}. See also Ref. \cite{JKG}.

\bibitem{ICTP} A. Bottino and N. Fornengo, Lectures given at the
  Fifth School on {\em Non-Accelerator Particle Astrophysics}, Abdus
  Salam International Centre for Theoretical Physics, Trieste, June
  1998, ed. G. Giacomelli, N. Paver and R.A.  Carrigan
  (hep-ph/hep-ph/9904469)
  
\bibitem{JKG} G. Jungman, M. Kamionkowski, K. Griest
  Phys. Rep. {\bf 267}, 195 (1996)
  
\bibitem{old} See, for instance: A. Bottino, V. de Alfaro, N.
  Fornengo, S. Mignola, S.  Scopel, Astropart.Phys {\bf 2}, 77 (1994);
  V. Berezinsky et al., Astropart.Phys. {\bf 5}, 1 (1996);
  L. Bergstrom and P. Gondolo, Astropart.Phys. {\bf 5}, 263 (1996); 
  V.A.Bednyakov, S.G.Kovalenko, H.V.Klapdor-Kleingrothaus,
  Y.Ramachers, Z.Phys. {\bf A357}, 339 (1997)

\bibitem{probing} A.Bottino, F. Donato, N.Fornengo and S. Scopel,
   hep-ph/0010203


\bibitem{further} A. Bottino, F. Donato, N. Fornengo, S. Scopel,
  Phys.Rev. {\bf D62}, 056006 (2000) 

\bibitem{notadir} For a complete updated list of references on recent papers
  on direct detection, see Ref. \cite{probing}


\bibitem{GPS} W.H. Press and D.N. Spergel, Ap. J {\bf 294}, 663
  (1985); Ap. J {\bf 296}, 679 (1985); A. Gould, Ap. J. {\bf 321}, 571
  (1987); A. Gould, Ap. J. {\bf 328}, 919 (1988); A. Gould, Ap. J.
  {\bf 368}, 610 (1991); A. Gould, Ap. J. {\bf 388}, 338 (1992)


\bibitem{noi_nuflux} A.Bottino, N.Fornengo, G.Mignola and L.Moscoso,
  Astropart. Phys. {\bf 3}, 65 (1995) 65
  
\bibitem{altri_nuflux} L. Bergstr\"om, J. Edsj\"o and P. Gondolo,
  Phys. Rev. D {\bf 58}, 103519 (1998).

  
\bibitem{nusugra} V. Berezinsky, A. Bottino, J. Ellis, N. Fornengo,
   G. Mignola, and S. Scopel, Astropart. Phys. {\bf 5}, 333 (1996).

\bibitem{oscill_exp} T. Montaruli (MACRO Collaboration), Proceedings
  of DARK98, Heidelberg, July 1998, hep-ex/9810017, A. Habig
  (Super--Kamiokande Collaboration), Proceedings of DPF'99,
  hep-ex/9903047

\bibitem{MACRO} The MACRO Collaboration, M. Ambrosio et al., Phys.Rev.
  D{\bf 60}, 082002 (1999) (hep-ex/9812020); G. Giacomelli and A.
  Margiotta, invited paper at the Chacaltaya Meeting on Cosmic Ray
  Physics, La Paz, July 2000 hep-ex/0010055
  
\bibitem{SK_WIMP} Super-Kamiokande Collaboration, contributed paper to
  30th International Conference on High Energy Physics,
  astro-ph/0007003

\bibitem{FGGV} See, for instance: 
N. Fornengo, M.C. Gonzalez-Garcia and J.W.F. Valle,
Nucl.Phys. {\bf B580} (2000) 58

\bibitem{SK_OSC_nos} Super-Kamiokande Collaboration, hep-ex/0009001

  
\bibitem{Ellis} The possibility of neutrino oscillation on the
  indirect signal from neutralino annihilation has also been
  considered in J. Ellis, R.A. Flores and S.S. Masood, Phys. Lett.
  {\bf B294}, 229 (1992) and A. de Gouv\^ea, hep-ph/0006157 in the
  case of the signal from the Sun and in M. Kowalski, preprint DESY
  00-125, hep-ph/0009183, for both the signals from the Earth and from
  the Sun

\bibitem{Kim} See, for instance: C.W. Kim and A. Pevsner, {\em
    Neutrinos in Physics and Astrophysics}, Contemporary Concepts in
  Physics, vol. 8 (Harwood Academic Press, Chur, switzerland, 1993)
  
\bibitem{meprev} N. Fornengo, Proceedings of WIN99, Cape Town, South
  Africa, January 1999 and of the EU Meeting 'Physics Beyond the
  Standard Model', SISSA, Trieste, February 1999, hep-ph/9904351


\end{thebibliography}
\end{document}